\documentstyle[psfig]{article}
\newcommand\etal{et al. }

\newcommand\ros{{\it Rosat\/}}

\newcommand\nh{N_{H}}

\newcommand\noteaa{1}
\newcommand\noteab{26}
\newcommand\noteac{2}
\newcommand\notead{3}
\newcommand\noteae{4}
\newcommand\noteaf{5}
\newcommand\noteag{6}
\newcommand\noteah{7}
\newcommand\noteai{8}
\newcommand\noteaj{9}
\newcommand\noteak{11}
\newcommand\noteal{15}
\newcommand\noteam{12}
\newcommand\notean{13}
\newcommand\noteao{14}
\newcommand\noteap{16}
\newcommand\noteaq{17}
\newcommand\notear{18}
\newcommand\noteat{19}
\newcommand\noteau{20}
\newcommand\noteav{21}
\newcommand\noteaw{22}
\newcommand\noteax{23}
\newcommand\noteay{24}
\newcommand\noteaz{27}
\newcommand\noteba{28}
\newcommand\notebb{29}
\newcommand\notebc{30}
\newcommand\notebd{31}
\newcommand\notebe{32}
\newcommand\notebf{33}
\newcommand\notebg{25}
\newcommand\notebh{10}
\newcommand\notebi{34}

\begin{document}

{\Large \bf 
\noindent
{Submillimeter Evidence for the Coeval Growth of Massive Black Holes and
Galaxy Bulges
}}
\vspace{4mm}

{\bf
\noindent
M.J. Page\(^{1}\), J.A. Stevens\(^{1}\), J.P.D. Mittaz\(^{1}\), 
F.J. Carrera\(^{2}\)
}
\vspace{2mm}

\noindent
\(^{1}\)Mullard Space Science Laboratory, University College London,
Holmbury St Mary, Dorking, Surrey RH5 6NT, UK.

\noindent
\(^{2}\)Instituto de F\'\i sica de Cantabria (Consejo Superior de
Investigaciones Cient\'\i ficas--Universidad de Cantabria), 39005
Santander, Spain.
\vspace{4mm}

{\bf
\noindent

The correlation, found in nearby galaxies, 
between black hole mass and stellar bulge mass 
implies that the formation of these
two components must be related.
Here we report submillimeter
photometry of eight x--ray absorbed active galactic
nuclei which have luminosities and redshifts 
characteristic of the sources that produce the bulk of the 
accretion luminosity
in the universe. 
The four sources with the highest redshifts
are
detected 
at 850 microns, with flux densities between
5.9 and 10.1 milliJanskies, 
and hence are ultraluminous infrared galaxies.  Interpreting
the submillimeter flux as emission from dust heated by starbursts, these
results suggest that the majority of stars in spheroids were formed at the same
time as their central black holes built up most of their mass by 
accretion, 
accounting for
the observed demography of massive black holes in the local universe. The
skewed rate of submillimeter detection with redshift is consistent with a high
redshift epoch of star formation in radio quiet active galactic nuclei, similar
to that seen in radio galaxies.  }

\vspace{2mm}

\noindent

In the local universe, central black holes are
found in most galaxy spheroids (a collective term for elliptical galaxies and 
the bulges of spiral galaxies)
with mass roughly proportional to that
of the spheroid (0.13\% $\pm 0.4$ dex) ({\it \noteaa}). The
simplest explanation for this proportionality is that the black hole
mass is built up in active galactic nuclei (AGN) by accretion of the same gas 
that is rapidly forming
the stars which make up the spheroid, i.e. 
the formation of the spheroid and the growth of the massive
black hole are coeval. Assuming that 10\% of the spheroid
mass $M$ is converted from hydrogen to helium in stars ({\it \noteac}) 
at an efficiency of 0.007 and radiated, and that 0.13\% of the spheroid
mass $M$ is accreted by the central black hole ({\it \noteaa}) at 10\% efficiency, the
ratio of radiation emitted by the starburst ($E_{SB}$) to that emitted
by the AGN ($E_{AGN}$)is:
\begin{equation}
\frac{E_{SB}}{E_{AGN}}=\frac{0.1 M \times 0.007}{0.0013 M \times 0.1}\sim 5
\label{eq:spheroidblackhole}
\end{equation}
and hence if they have similar lifetimes 
we expect the stellar component to have around 5 times the bolometric
luminosity of the AGN
during the spheroid formation;
a similar argument
has been used to estimate
the relative contribution of accretion and nuclear fusion to the
extragalactic background radiation ({\it \notead}). 
The
scatter on this relation for individual objects is expected to be at least 
0.4 dex to 
account for the
scatter on the present day spheroid/black hole mass ratio ({\it \noteaa}), 
with additional scatter of $\sim$ 0.3 dex from the
intrinsic variability of each AGN ({\it \noteae}).

To investigate this hypothesis, 
we have studied a sample of eight $z>1$, x--ray absorbed AGN 
discovered
serendipitously in archival x--ray 
data
(Table 1). 
They were chosen from the 
fourteen such sources reported ({\it \noteaf}) on the basis of visibility 
in our allocated observing shifts at the James Clerk Maxwell Telescope (JCMT). 
If the stellar and black hole components of present day galaxy 
spheroids did form
at the same time, then the 
majority of star formation in spheroids must have taken place around the AGN 
that dominate the universe's accretion power, i.e. those that are
responsible for the majority of present-day black hole mass. 
Our targets are representative of these AGN for four reasons.
Firstly, their redshifts span the $1<z<3$ epoch in which the accretion
luminosity from AGN peaked.  For example in the best fit luminosity function 
of a large sample of x--ray selected AGN ({\it \noteag}), 
the comoving AGN x--ray luminosity density is more than a
factor 10 larger at $z=2$ than in the local universe.
Secondly, their luminosities are close to the break in the luminosity
function, which is where the majority of AGN accretion luminosity is
produced.  For example in the same 
model x--ray
luminosity function ({\it \noteag}), 60\% of the comoving AGN
x--ray luminosity density is produced by AGN within $\pm 0.5$ dex of
the break in the luminosity function, which is at $\log L_{X} = 44.5$
at $z=2$.
Thirdly, they are x--ray absorbed sources. It has been estimated 
({\it \noteah}) that
85\% of accretion power in the universe is absorbed, and
x--ray background synthesis models require that the majority of AGN
are intrinsically absorbed ({\it \noteai}).
Finally, most of them (all but one) are radio quiet; close to the break in the
luminosity function, radio quiet AGN outnumber radio loud AGN by about
15:1 ({\it \noteaj}).
\vspace{1mm}

Observations at 850$\mu$m were carried out at the JCMT in excellent,
stable weather conditions ({\it \notebh}) on the 18th -- 20th January 2001 using the
Submillimeter Common User Bolometer Array (SCUBA) ({\it \noteak}). 
Four of the eight sources have
significant detections ($> 5 \sigma$) at 850$\mu$m with flux densities
between 5.9 and 10.1 mJy; the other four
sources were not detected (Table 1).
The high submillimeter detection rate for our sources (50\%)
contrasts 
with the low rates of detection ($\leq$ 10\%) in 
previous submillimeter surveys 
of x--ray selected objects ({\it \noteam}--{\it \noteap}). 
However, the x--ray sources investigated here have significantly higher 
x--ray flux (by more than a factor of 3) than the relatively 
faint 
x--ray sources in the other surveys, and hence 
the faint x--ray sources would probably not be
detectable at 850$\mu$m even if they had the same ratio of
submillimeter to x--ray flux as our 850 $\mu$m detected sources. 
Although our sample is small, our high detection rate means that at the
99\% confidence level $>$ 12\% ({\it \noteap}) 
of the population from which our 
sample was drawn are 850$\mu$m sources detectable with SCUBA; at the 95\% 
confidence 
level this figure rises to $>$ 19\%. 

We interpret the observed
submillimeter flux from all our sources as thermal emission from dust.
Because RXJ163308.57+570258.7 is radio loud ({\it \noteaf}), we have also considered 
(but rejected) the possibility that
synchrotron emission contributes to the flux from this source at 
submillimeter wavelengths. 
The
source has a 1.4 GHz flux density of 17.2$\pm 0.7$ mJy in the northern Very
Large Array sky
survey ({\it \noteaq}) and a 325 MHz flux density of 78$\pm 4$ mJy
in the Westerbork northern sky survey ({\it \notear}). It is
therefore a steep spectrum source and extrapolating the
spectrum as a power law we expect it to contribute an insignificant amount ($<0.1$ mJy) at
850$\mu$m. 
We therefore conclude that
the observed 850$\mu$m flux of RXJ163308.57+570258.7 is 
dominated by thermal emission from dust.
For all our sources, we have estimated the total far infrared (FIR) 
luminosities $L_{FIR}$ and dust masses $M_{d}$ (Table
\ref{tab:results}) from the monochromatic 850$\mu$m fluxes 
using the FIR spectrum of Mrk~231 as a template. 
Mrk~231 is appropriate because it is a well
studied, nearby ultraluminous infrared galaxy (ULIRG), which is 
similar to our AGN in that it hosts an x--ray absorbed,
broad line, active nucleus, and has 
similar submillimeter luminosity (see Fig. 1).  
We have fitted an
isothermal optically thin dust model ({\it \noteat}) to the FIR data
on Mrk 231 ({\it \noteau})
and obtained a good fit for dust temperature $T= 44$~K, $\beta$=1.55
(where $\beta$ is the power law index of the frequency dependence of
the dust emissivity) and FIR luminosity of
$3.9\times 10^{12}$ L$_{\odot}$. 
The FIR luminosities of the four detected AGN qualify them 
as ULIRGs, and the FIR luminosity of RXJ094144.51+385434.8
in particular is sufficient for it to be classed as a `hyperluminous' infrared
galaxy ({\it \noteav}). 
The current upper
limits to the submillimeter fluxes of the four undetected AGN are 
larger than the flux expected from Mrk 231 at equivalent
redshifts (Fig. 1), and it is therefore possible that the entire sample 
could be ULIRGs.

An important question is the relative contribution of 
AGN-heated and starburst-heated dust to the FIR luminosities of 
ULIRGs. 
Recent results
from {\em ISO\/} mid-infrared spectroscopy ({\it \noteaw}), 
millimeter interferometry ({\it \noteax}) and detailed radiative transfer
modeling ({\it \noteav}) suggest that massive stars are the
dominant power source of around three quarters of nearby ULIRGs 
(including Mrk~231) and hyperluminous infrared galaxies.
For our four sources detected at
850$\mu$m, comparing the power output of the AGN to the power emitted
in the FIR gives an indication of the relative importance of AGN and
starburst heating of the FIR emitting dust.
Assuming that 3\% of the bolometric luminosity of an AGN 
is emitted in the 0.5 - 2
keV band ({\it \noteay},{\it \notebg}), 
the bolometric luminosities of the AGN in
RXJ094144.51+385434.8 and RXJ121803.82+470854.6 are about one quarter and one
third of their FIR luminosities respectively, while the AGN and FIR
luminosities are equivalent (within 0.1 dex) in RXJ124913.86-055906.2 and
RXJ163308.57+570258.7. 
This means that in RXJ094144.51+385434.8 and 
RXJ121803.82+470854.6 the AGN are not sufficiently powerful to heat
the FIR emitting dust, while in the other two cases all 
of the AGN radiation
would have to be absorbed and re-emitted by dust for the AGN to power the 
FIR emission.
The low ratios of AGN to FIR luminosity therefore make it very
likely that in these sources
the
FIR is predominantly powered by starlight rather than the central AGN.
Our submillimeter detections thus imply that these sources contain massive
reservoirs of molecular gas and are
producing stars at a prodigious rate ($> 1000$ $M_{\odot}$ yr$^{-1}$).

These results have important implications for our understanding of the
formation of galaxy spheroids and
massive black holes. 
 The four AGN
we have detected at 850$\mu$m all have FIR and AGN luminosities similar 
to what would be expected from equation \ref{eq:spheroidblackhole}
(this could also be true for the four sources
undetected at 850$\mu$m: this possibility is not excluded by the current 
850$\mu$m
upper limits). Our 850$\mu$m detection rate thus implies that with 95\% 
confidence, between 19\% and 100\% of the the population from which 
our sample is drawn 
are simultaneously building up their black hole mass by accretion and 
undergoing intense star formation at rates which are  
consistent with, indeed suggestive of, 
the spheroid/AGN formation scenario outlined at the beginning of this letter.

The
redshift distribution of our detections is bimodal, in that all the
sources with $z>1.5$ have been detected at 850$\mu$m compared to none
of the sources with $z<1.5$. This is
unlikely to be coincidence: if we split the sample in two by redshift,
and assume that the underlying probability of submillimeter detection
were independent of redshift, then the probability of detecting all
the sources in one redshift bin and detecting none of the sources in
the other is $ < 1\%$. However, the two highest redshift objects in the 
sample also have the highest x--ray luminosities, and hence a correlation 
between x--ray and submillimeter luminosity could be responsible for the 
bimodal detection rate. Nonetheless, the skewed detection rate is in the same 
sense as the strong dependence of submillimeter luminosity with redshift that 
has already been found in a sample of radio galaxies ({\it \noteab}),
suggesting that radio galaxies had higher rates of star formation at 
earlier epochs. 
Age determinations of stellar bulges in nearby luminous AGN suggest that 
radio quiet AGN, as well as radio loud AGN,
had higher rates of star formation at earlier epochs ({\it \noteaz}); 
the trend seen in our data is 
consistent 
with this hypothesis.

A consequence of coeval spheroid and black hole formation for the
recently discovered luminous submillimeter population ({\it \noteba}), would
be that the majority must host AGN. Indeed, it is already known
that a significant fraction of the most luminous submillimeter
galaxies found so far in deep SCUBA surveys contain AGN ({\it \notebb}).  At
redshifts of 2 -- 3, starbursts with similar bolometric luminosity to
AGN with $44 < \log L_{X} < 45$ would have 850$\mu$m fluxes of $\sim$
0.5 -- 5 mJy, and this is the flux range in which the
bulk of the cosmic 850$\mu$m background is produced ({\it \notebc}).  Current
x--ray luminosity functions ({\it \noteag},{\it \notebd}) suggest that there 
are at least
several hundreds of AGN deg$^{-2}$ with $z>1$ and $\log L_{X} > 44$
assuming a 4:1 ratio of obscured to unobscured sources. Although this
falls well short of the current 850$\mu$m source counts (8000$\pm$3000
deg$^{-2}$ at 1 mJy) ({\it \notebc}), the AGN x--ray luminosity function has a
large uncertainty at $z>2$, which is the epoch in which the
majority of luminous submillimeter galaxies are
found ({\it \notebe}). Furthermore, the space density of obscured AGN at high
redshift is unknown: the x--ray background intensity does
not preclude the existence of a large population of high redshift
Compton-thick sources.  

\section{References and notes}
\small

\noindent \noteaa. J. Kormendy and K. Gebhardt, in {\it The 20th Texas Symposium on
Relativistic Astrophysics}, H. Martel and J.C. Wheeler, Eds, AIP, in press 
(astro-ph/0105230)

\noindent \noteac. M. Sch\"onberg, and S. Chandrasekhar, 
{\it Astrophys. J.} 96, 161 (1942).

\noindent \notead. G. Hasinger, 
in {\it ISO Surveys of a Dusty Universe}, D. Lemke, M. Stickel, K. Wilke, Eds.,
Springer, in press (astro-ph/0001360)

\noindent \noteae. M.R.S. Hawkins, 
{\it Astron. and Astrophys. Suppl.} 143, 465 (2000).

\noindent \noteaf. M.J. Page, J.P.D. Mittaz, F.J. Carrera, 
{\it Mon. Not. R. Astron. Soc.} 325, 575 (2001).

\noindent \noteag. M.J. Page, K.O. Mason, I.M. McHardy, L.R. Jones, F.J. Carrera,
{\it Mon. Not. R. Astron. Soc.} 291, 324 (1997).

\noindent \noteah. A.C. Fabian and K. Iwasawa, 
{\it Mon. Not. R. Astron. Soc.} 303, L34 (1999).

\noindent \noteai. R. Gilli, G. Risaliti, M. Salvati, 
{\it Astron. and Astrophys.} 347, 424 (1999).

\noindent \noteaj. P. Ciliegi, 
M. Elvis, B.J. Wilkes, B.J. Boyle, R.G. McMahon, T. Maccacaro, 
{\it Mon. Not. R. Astron. Soc.} 277, 1463 (1995).

\noindent \notebh. 
Observations were carried out in photometry mode using a standard
chop/nod/jiggle observing technique.  The atmospheric
transmission and `submillimeter seeing' were at all times in the top
quartile of values measured on Mauna Kea. STARLINK {\small SURF} software
was used to
correct for the nod, flatfield, extinction correct, despike and remove sky
noise from the data.  Flux calibration was made against the primary
submillimeter calibrator, Mars.

\noindent \noteak. W.S. Holland, \etal 
{\it Mon. Not. R. Astron. Soc.} 303, 659 (1999).

\noindent \noteam. A.E. Hornschemeier, \etal
{\it Astrophys. J.} 554, 742 (2001).

\noindent \notean. A.J. Barger, L.L. Cowie, R.F. Mushotzky, U.A. Richards, 
2001, {\it Astron. J.} submitted, (astro-ph/0007175)

\noindent \noteao. P. Severgnini, \etal 
{\it Astron. and Astrophys.} 360, 457 (2000).

\noindent \noteal. A.C. Fabian, \etal
{\it Mon. Not. R. Astron. Soc.} 315, L8 (2000).

\noindent \noteap. N. Gehrels, 
{\it Astrophys. J.} 303, 336 (1986).

\noindent \noteaq. J.J. Condon, \etal
{\it Astron. J.} 115, 1693 (1998).

\noindent \notear. R.B. Rengelink, \etal
{\it Astron. and Astrophys. Suppl.} 124, 259 (1997).

\noindent \noteat. R.D. Hildebrand, 
{\it Quarterly J. R. Astron. Soc.} 24, 267 (1983).

\noindent \noteau. D. Rigopoulou, A. Lawrence, M. Rowan-Robinson, 
{\it Mon. Not. R. Astron. Soc.} 278, 1049 (1996).

\noindent \noteav. M. Rowan-Robinson, 
{\it Mon. Not. R. Astron. Soc.} 316, 885 (2000).

\noindent \noteaw. R. Genzel, \etal, 
{\it Astrophys. J.} 498, 579 (1998).

\noindent \noteax. D. Downes, P.M. Solomon, 
{\it Astrophys. J.} 507, 615 (1998).

\noindent \noteay. M. Elvis, \etal,
{\it Astrophys. J. Suppl.} 95, 1 (1994).

\noindent \notebg. If our objects are systematically underluminous in
x--rays then we will be underestimating the bolometric luminosities of 
their active nuclei. However, this is unlikely because 
these sources were found in an x-ray survey, in which 
the natural selection bias 
is expected to favour 
sources with higher than average x-ray to bolometric 
luminosity ratios rather than the reverse. 

\noindent \noteab. 
E.N. Archibald, J.S. Dunlop, D.H. Hughes, S. Rawlings, S.A. Eales, 
R.J. Ivison, 
 {\it Mon. Not. R. Astron. Soc.} 323, 417 (2001).

\noindent \noteaz. L.A. Nolan, J.S. Dunlop, M.J. Kukula, D.H. Hughes, T. Boroson, 
R. Jimenez,
{\it Mon. Not. R. Astron. Soc.} 323, 308 (2001).

\noindent \noteba. I. Smail, R.J. Ivison, A.W. Blain,
{\it Astrophys. J.} 490, L5 (1997)

\noindent \notebb. R.J. Ivison, \etal,
{\it Mon. Not. R. Astron. Soc.} 315, 209 (2000).

\noindent \notebc. A.W. Blain, J.-P. Kneib, R.J. Ivison, I. Smail, 
{\it Astrophys. J.} 512, L87 (1999).

\noindent \notebd. T. Miyaji, G. Hasinger, M. Schmidt, 
{\it Astron. and Astrophys.} 369, 49 (2001).

\noindent \notebe. J.S. Dunlop, 
 in {\it FIRSED2000}, I.M. van Bemmel, B. Wilkes, P. Barthel, Eds 
conference proceedings, {\it Elsevier New Astron. Rev.}, in press,
(astro-ph/0101297).

\noindent \notebf. B.T. Draine and H.M. Lee, 
{\it Astrophys. J.} 285, 89, (1984).

\noindent \notebi. 
The James Clerk Maxwell Telescope is operated on behalf of the Particle Physics
and Astronomy Research Council of the United Kingdom, the Netherlands
Organization for Scientific Research and the National Research Council of
Canada.

\newpage

\begin{table*}
\caption{Characteristics of the observed x--ray absorbed AGN and their 
observed submillimeter emission. We assume a Hubble constant $H_{0}$ = 50 km
s$^{-1}$ Mpc$^{-1}$, a deceleration parameter $q_{0}$ = 0.5 and zero
cosmological constant. For sources that were not detected with SCUBA we have
estimated upper limits to $L_{FIR}$ and $M_{d}$ by taking the upper
limit of $S_{850}$ to be $S_{850}$ +2$\sigma$ for sources with positive
$S_{850}$, and 2$\sigma$ for sources with negative $S_{850}$.
}
\label{tab:results}
\scriptsize
\begin{tabular}{lccccccccc}
1&2&3&4&5&6&7&8&9\\
Source&$z$&$S_{X}$&log $L_{X}$&log $\nh$ &RL/RQ& 
$S_{850}$  & Dust Mass &$L_{FIR}$ \\
                & & &
&     
&   &   (mJy)  &     ($M_{\odot}$)   & ($L_{\odot}$)   \\
&&&&&&&&\\
RXJ094144.51+385434.8&1.819&$2.1^{+0.5}_{-0.5}$&$44.8\pm0.1$&$21.9^{+0.5}_{-0.4}$&RQ&
$10.1\pm1.7$& $1.2 \times 10^{9}$ & $2.0\times 10^{13}$ \\
RXJ101123.17+524912.4&1.012&$3.3^{+1.0}_{-0.9}$&$44.7\pm0.2$&$22.5^{+0.2}_{-0.3}$&RQ
&$-1.4\pm1.9$& $<4.7 \times 10^{8}$ & $<8.0\times 10^{12}$ \\
RXJ104723.37+540412.6&1.500&$1.7^{+0.7}_{-0.6}$&$44.6\pm0.2$&$22.2^{+0.4}_{-0.6}$&RQ
&$2.3\pm1.8$ &   $<7.2 \times 10^{8}$ & $<1.2\times 10^{13}$        \\
RXJ111942.16+211518.1&1.288&$3.4^{+0.6}_{-0.5}$&$44.6\pm0.1$&$21.4^{+0.4}_{-0.3}$&RQ
&$-0.9\pm1.8$&   $<4.5 \times 10^{8}$ & $<7.6\times 10^{12}$        \\
RXJ121803.82+470854.6&1.743&$1.5^{+0.4}_{-0.4}$&$44.7\pm0.2$&$22.3^{+0.3}_{-0.7}$&RQ
&$6.8\pm1.2$ &   $8.0 \times 10^{8}$ & $1.4\times 10^{13}$        \\
RXJ124913.86$-$055906.2&2.212&$2.4^{+0.5}_{-0.4}$&$45.1\pm0.1$&$22.2^{+0.4}_{-0.6}$&RQ
&$7.2\pm1.4$ &   $7.8 \times 10^{8}$ & $1.3\times 10^{13}$        \\
RXJ135529.59+182413.6&1.196&$3.6^{+0.9}_{-0.9}$&$44.7\pm0.2$&$22.2^{+0.2}_{-0.5}$&RQ
&$0.8\pm1.3$ &   $<4.3 \times 10^{8}$ & $<7.2\times 10^{12}$        \\
RXJ163308.57+570258.7&2.802&$1.9^{+0.3}_{-0.3}$&$45.2\pm0.1$&$22.5^{+0.3}_{-0.5}$&RL
&$5.9\pm1.1$ &   $5.8 \times 10^{8}$ & $9.8\times 10^{12}$        \\
&&&&&&&&\\
\multicolumn{9}{l}{1 Source name based on \ros\ position. RXJ124913.86$-$055906.2 
is also known as [HB89] 1246$-$057,}\\
\multicolumn{9}{l}{\ \ \ \ 
RXJ135529.59+182413.6 is also known as RIXOS F268\_011, and 
RXJ163308.57+570258.7 is also known as WN B1632+5709}\\
\multicolumn{9}{l}{2 Redshift}\\
\multicolumn{9}{l}{3 Observed 0.5 -- 2 keV flux 
(10$^{-14}$ erg cm$^{-2}$ s$^{-1}$)}\\
\multicolumn{9}{l}{4 Log [0.5 -- 2 keV luminosity (erg s$^{-1}$)] corrected 
for intrinsic absorption}\\
\multicolumn{9}{l}{5 Log [intrinsic column density (cm$^{-2}$)]}\\
\multicolumn{9}{l}{6 Radio loud (RL) if $\alpha_{OR} > 0.35$ or radio quiet
(RQ) if $\alpha_{OR} < 0.35$}\\
\multicolumn{9}{l}{7 850$\mu$m flux. The quoted errors do not include 
calibration uncertainties of 10\%.}\\
\multicolumn{9}{l}{8 Dust mass inferred from 850$\mu$m flux, 
adopting a value for the dust mass
absorption coefficient ({\it \notebf}) 
$\kappa_{100\mu m}$ = 5.5 m$^{2}$ kg$^{-1}$
}\\
\multicolumn{9}{l}{9 FIR luminosity inferred from 850$\mu$m flux}\\
\end{tabular}
\vspace{7cm}
\end{table*}

\newpage

\begin{figure}
\begin{center}
\leavevmode
\setlength{\unitlength}{1cm}
\begin{picture}(8.8,12.4)
\put(0,0){\includegraphics{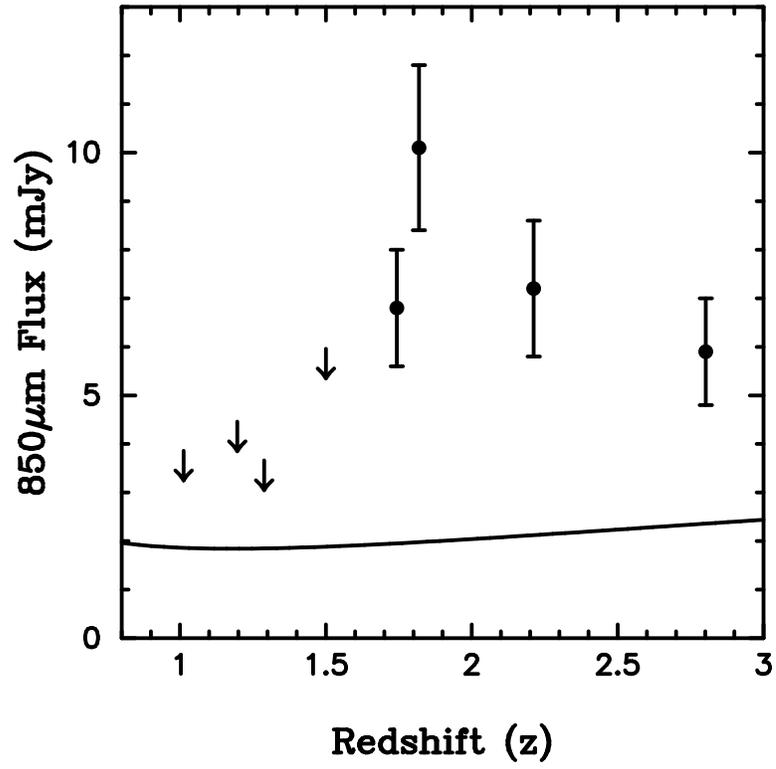}}
\end{picture}
\caption{850$\mu$m fluxes of the X-ray absorbed AGN as a function of 
redshift $z$.
The solid line shows the predicted 850$\mu$m flux of Mrk 231 if it were 
viewed at redshift $z$.}
\label{fig:hardflux}
\vspace{7cm}
\end{center}
\end{figure}

\end{document}